\documentstyle[aps,prd,epsf]{revtex}
\begin{document}
\draft
\title{Self-interacting Warm Dark Matter}
\author{Steen Hannestad}
\address{NORDITA, Blegdamsvej 17, DK-2100 Copenhagen, Denmark}
\author{Robert J.~Scherrer}
\address{Department of Physics and Department of Astronomy,
Ohio State University, Columbus, OH 43210}
\date{\today}

\maketitle

\begin{abstract}
It has been shown by many independent studies that the cold dark matter
scenario produces singular galactic dark halos, in strong contrast with
observations. Possible remedies are that either the dark matter is warm so that
it has significant thermal motion or that the dark matter has strong 
self interactions.  We have combined these ideas to calculate the
linear mass power spectrum and the spectrum of cosmic microwave background
(CMB) fluctuations for self-interacting warm dark matter.  Our results
indicate that such models have more power on small scales than
is the case for the standard warm dark matter model, with a CMB
fluctuation spectrum which is nearly indistinguishable
from standard cold dark matter.  This enhanced small-scale power
may provide better agreement with the observations than does standard
warm dark matter.
\end{abstract}



\section{introduction}

Dark matter is a necessary ingredient in the standard Big Bang model of the 
universe. Its presence has an impact from subgalactic dynamics to the
global evolution of the universe. However, the nature of the dark matter
remains unknown. So far, the cold dark matter model has been very successful
in explaining how structure forms \cite{peacock,gross}. 
In this model the dark matter consists
of weakly interacting massive particles (WIMPs) which are extremely non-relativistic
when structure formation begins. Because they are so massive they do not
free stream and perturbations on small scales are preserved. In the 1980s 
it was realised that CDM produces too much small-scale structure, 
and that some modification
of the model is needed. Several possibilities exist: there could be a large
component of hot dark matter damping small scale fluctuations or there
could be a non-zero cosmological constant. Recent data from type Ia supernovae indeed
suggest that the energy density of the universe is dominated by a cosmological
constant \cite{SNIa}. Thus, the problem with CDM is at first sight remedied.
However, in the past few years very high resolution N-body simulations of 
structure formation have shown that any type of CDM model produces far too
much substructure on galactic scales, compared with observations.
The halo of a galaxy like our own should contain of the order 1000 
distinct subhaloes, a factor of ten more than is found by observations
\cite{moore,ghigna}.
Another, related problem is that galaxies are predicted to have singular cores.
Navarro, Frenk and White \cite{NFW96} 
found that N-body simulations predicted a universal
core profile of halos where $\rho \propto r^{-1}$. Later simulations with
higher resolution find an even steeper profile \cite{FP94,N99,NS99,moore2}.
At the same time galactic rotation curves indicate dark matter halos with
finite cores, i.e. constant core density \cite{paolo}.
This problem is very severe and is consistently found in all simulations. 

If the details of star formation and feedback do not solve the problem,
then physics at a more fundamental level possibly could. 
One option is that the primordial power spectrum has a sharp drop at
subgalactic scales so that substructure is prevented from forming \cite{KL99}.
Another option along this line is that the dark matter is not cold, but warm
\cite{SS88,SLD99}.
In this model the dark matter particle mass should be around 1 keV so that
the dark matter has significant thermal motion and perturbations on
small scales are erased. However, the cut-off scale needed for the correct
core radius of halos to be produced is so large that it is difficult to form 
the correct number of dwarf galaxies \cite{HD}.

A radically different explanation was suggested by Spergel and Steinhardt \cite{SS99},
namely that the dark matter could be cold, but have significant self-interactions.
If the mean free path of the dark matter particles is of the order the size of the
collapsing system, then the core singularity would form much more slowly,
while the outer parts of the halo would remain unchanged.
Recently, a large number of papers have appeared which investigate this
possibility numerically \cite{H99,burkert,firmani,yoshida,MGJPQ,ostriker}. 
The conclusion is that if the interactions
are very strong, the model does not fit observations 
\cite{burkert,firmani,yoshida,MGJPQ,ME}.
The halos become completely
spherical apart from a small rotational deformation, and a singular core develops.
However, it seems that models where the dark matter mean free path is similar to the
system size produce halos closely resembling the observed ones
\cite{burkert,firmani}. It has also been suggested that the self-interacting
matter could
be in the form of a scalar field \cite{peebles}.

That dark matter could have self-interactions is an old idea. It was originally
suggested by Raffelt and Silk \cite{RS} 
that HDM neutrinos could have strong self interactions.
In this way free streaming would be suppressed and fluctuations only washed
out via diffusion. 
The scenario was elaborated on by Atrio-Barandela and
Davidson \cite{AD} who did a numerical study of this model.
The possibility of number changing self interactions has also been considered
\cite{CMH92,machacek,LSS}.

In the present paper we wish to explore the possibility that dark matter has
both significant thermal motion and self-interactions. 
The self-interactions are assumed to consist only of two-particle scattering.
In general, the inclusion
of self interactions leads to less small scale suppression of perturbations
because the small scale cut-off in power is given by the Jeans scale which is
smaller than the free-streaming scale.
We find that self-interacting hot dark matter, as suggested by Refs.~\cite{RS,AD},
is clearly ruled out because it produces far too little small-scale structure.
However, self interacting warm dark matter may be a viable possibility. Strong
self interactions push the power spectrum towards smaller scales by roughly a
factor of 1.6, which may make it consistent with observations.


\section{The Boltzmann equation}

The evolution of any given particle species can be described via the Boltzmann
equation. Our notation is identical to that of Ma and Bertschinger (MB) \cite{MB}.
We shall work in synchronous gauge because the numerical routine for calculating
matter and CMB power spectra, CMBFAST \cite{SZ96}, is written in this gauge.
As the time variable we use conformal time, defined as $d \tau = dt/a(t)$, where
$a(t)$ is the scale factor. Also, as the momentum variable we shall use the
comoving momentum $q_j \equiv a p_j$. We further parametrize $q_j$ as
$q_j = q n_j$, where $q$ is the magnitude of the comoving momentum and
$n_j$ is a unit 3-vector specifying direction.

The Boltzmann equation can generically be written as
\begin{equation}
L[f] = \frac{Df}{D\tau} = C[f],
\end{equation}
where $L[f]$ is the Liouville operator. 
The collision operator on the right-hand side describes
any possible collisional interactions.

We then write the distribution function as
\begin{equation}
f(x^i,q,n_j,\tau) = f_0(q) [1+\Psi(x^i,q,n_j,\tau)],
\end{equation}
where $f_0(q)$ is the unperturbed distribution function.
For a standard fermion which decouples while relativistic, this distribution
function is simply
\begin{equation}
f_0(q) = [\exp(q/T_0)+1]^{-1},
\end{equation}
where $T_0$ is the present-day temperature of the species. For a self-interacting
species in scattering equilibrium the distribution is instead
\begin{equation}
f_0(q) = [\exp((\epsilon-\mu)/aT)+1]^{-1},
\end{equation}
where $\epsilon = \sqrt{q^2+a^2 m^2}$ and $\mu$ is a chemical potential.
This distribution is in general different from the one for collisionless particles,
so that one might worry that a detailed calculation of $f_0(q,\tau)$ is needed.
However, the relevant quantity to look at for our purpose is the entropy per
particle, $s/n$, which is conserved for both interacting and
non-interacting species
(note that this would not hold in a model with number-changing self
interactions \cite{CMH92,machacek,LSS}). 
This means that for instance $\langle p/T_\gamma \rangle
= {\rm constant}$. Thus we do not need to worry about how the unperturbed distribution
is changed by self-interactions.
In practise we just assume that the distribution function is equal to what it
would be for a collisionless species.

In synchronous gauge the Boltzmann equation can be written as an evolution
equation for $\Psi$ in $k$-space \cite{MB}
\begin{equation}
\frac{1}{f_0} L[f] = \frac{\partial \Psi}{\partial \tau} + i \frac{q}{\epsilon}
\mu \Psi + \frac{d \ln f_0}{d \ln q} \left[\dot{\eta}-\frac{\dot{h}+6\dot{\eta}}
{2} \mu^2 \right] = \frac{1}{f_0} C[f],
\end{equation}
where $\mu \equiv n^j \hat{k}_j$.
$h$ and $\eta$ are the metric perturbations, defined from the perturbed space-time
metric in synchronous gauge \cite{MB}
\begin{equation}
ds^2 = a^2(\tau) [-d\tau^2 + (\delta_{ij} + h_{ij})dx^i dx^j],
\end{equation}
\begin{equation}
h_{ij} = \int d^3 k e^{i \vec{k}\cdot\vec{x}}\left(\hat{k}_i \hat{k}_j h(\vec{k},\tau)
+(\hat{k}_i \hat{k}_j - \frac{1}{3} \delta_{ij}) 6 \eta (\vec{k},\tau) \right).
\end{equation}

{\it Collisionless Boltzmann equation ---}
At first we assume that $\frac{1}{f_0} C[f] = 0$.
We then expand the perturbation as 
\begin{equation}
\Psi = \sum_{l=0}^{\infty}(-i)^l(2l+1)\Psi_l P_l(\mu).
\end{equation}
One can then write the collisionless
Boltzmann equation as a moment hierarchy for the $\Psi_l$
by performing the angular integration of $L[f]$
\begin{eqnarray}
\dot\Psi_0 & = & -k \frac{q}{\epsilon} \Psi_1 + \frac{1}{6} \dot{h} \frac{d \ln f_0}
{d \ln q} \label{eq:psi0}\\
\dot\Psi_1 & = & k \frac{q}{3 \epsilon}(\Psi_0 - 2 \Psi_2) \label{eq:psi1}\\
\dot\Psi_2 & = & k \frac{q}{5 \epsilon}(2 \Psi_1 - 3 \Psi_3) - \left(\frac{1}{15}
\dot{h}+\frac{2}{5}\dot\eta\right)\frac{d \ln f_0}{d \ln q} \\
\dot\Psi_l & = & k \frac{q}{(2l+1)\epsilon}(l \Psi_{l-1} - (l+1)\Psi_{l+1}) 
\,\,\, , \,\,\, l \geq 3
\end{eqnarray}
It should be noted here that the first two hierarchy equations are directly
related to the energy-momentum conservation equation.
This can be seen in the following way. Let us define the density and
pressure perturbations of the dark matter fluid as \cite{MB}
\begin{eqnarray}
\delta & \equiv & \delta \rho/\rho \\
\theta & \equiv & i k_j \delta T^0_j/(\rho+P) \\
\sigma & \equiv & -(\hat{k}_i \hat{k}_j - \frac{1}{3} \delta_{ij})
(T^{ij}-\delta^{ij}T^k_k/3).
\end{eqnarray}
Then energy and momentum conservation implies that \cite{MB}
\begin{eqnarray}
\dot\delta & = & -(1+\omega)\left(\theta+\frac{\dot h}{2}\right)-
3 \frac{\dot a}{a} \left(\frac{\delta P}{\delta \rho} - \omega \right) \delta 
\label{eq:energy}\\
\dot \theta & = & \frac{\dot a}{a} (1-3 \omega)\theta - \frac{\dot \omega}{1+\omega}
\theta + \frac{\delta P/ \delta \rho}{1+\omega} k^2 \delta - k^2 \sigma.
\label{eq:mom}
\end{eqnarray}
By integrating Eq.~(\ref{eq:psi0}) over $q^2 \epsilon dq$,
one gets Eq.~(\ref{eq:energy}) and by integrating Eq.~(\ref{eq:psi1})
equation over $q^3 dq$ one retrieves Eq.~(\ref{eq:mom}).

{\it Collisional Boltzmann equation ---}
We now introduce interactions by lifting the restriction that 
$\frac{1}{f_0} C[f] = 0$. 
Ideally, one should calculate
the collision integrals in detail for some explicit interaction.
However, we shall instead use the cruder, but more model independent 
relaxation time approximation. Here, the right hand side of the
Boltzmann equation is in general written as \cite{relaxation}
\begin{equation}
\frac{1}{f_0} C[f] = -\frac{\Psi}{\tau},
\end{equation}
where $\tau$ is the mean time between collisions.
However, in this simple approximation we run the risk of not obeying
the basic conservation laws.
The collision term in Eq.~(\ref{eq:psi0}) is 
$\int d \Omega \frac{1}{f_0} C[f]$ and the one in 
Eq.~(\ref{eq:psi1}) is $\int d \Omega \mu \frac{1}{f_0} C[f]$.
Integrating these two terms over momentum space one gets the 
collision terms in Eqs.~(\ref{eq:energy}-\ref{eq:mom}) to be
\begin{equation}
\int C[f] d \Omega q^2 dq \epsilon
\end{equation}
and 
\begin{equation}
\int C[f] d \Omega q^2 dq \mu q = k^i \int C[f] d \Omega q^2 dq q_i
\end{equation}
respectively.
However, any integral of the form
\begin{equation}
\int C[f] d \Omega q^2 dq A,
\end{equation}
where $A \in (I,\epsilon,q_i)$ is automatically zero because $A$
is a collisional invariant (however, conservation of particle number ($I$)
only applies
to $2 \leftrightarrow 2$ scatterings). Thus, both the above integrals are
zero, and the right hand side of the $l=0$ and 1 terms should be zero,
reflecting that energy and momentum is conserved in each interaction. Apart from these
two terms we put
\begin{equation}
\frac{1}{f_0} C[f]_{l \geq 2} = -\frac{\Psi_l}{\tau},
\end{equation}
so that the full Boltzmann hierarchy, including interactions, is
\begin{eqnarray}
\dot\Psi_0 & = & -k \frac{q}{\epsilon} \Psi_1 + \frac{1}{6} \dot{h} \frac{d \ln f_0}
{d \ln q} \\
\dot\Psi_1 & = & k \frac{q}{3 \epsilon}(\Psi_0 - 2 \Psi_2) \\
\dot\Psi_2 & = & k \frac{q}{5 \epsilon}(2 \Psi_1 - 3 \Psi_3) - \left(\frac{1}{15}
\dot{h}+\frac{2}{5}\dot\eta\right)\frac{d \ln f_0}{d \ln q}
-\frac{\Psi_2}{\tau} \\
\dot\Psi_l & = & k \frac{q}{(2l+1)\epsilon}(l \Psi_{l-1} - (l+1)\Psi_{l+1})
-\frac{\Psi_l}{\tau} \,\,\, , \,\,\, l \geq 3
\end{eqnarray}
In Appendix A we discuss how the above set of equations relates to 
the equations used in other studies of self-interacting dark matter.

{\it Relaxation time ---}
We now need an expression for the collision time $\tau$. 
In general we can write
\begin{equation}
\tau^{-1} = n \langle \sigma |v| \rangle.
\end{equation}
For relativistic
particles scattering via exchange of a massive vector boson ($m_X \gg T,m$ where
$m_X$ is the vector boson mass and $m$ is the mass of the dark matter particle)
we have
\begin{equation}
\langle \sigma |v| \rangle \propto (T/m)^2,
\end{equation}
whereas for non-relativistic particles it is
\begin{equation}
\langle \sigma |v| \rangle \propto (T/m)^{1/2}.
\end{equation}
As an interpolation we use
\begin{equation}
\langle \sigma |v| \rangle = \frac{1}{2} \sigma_0 \left[ \left(\frac{T}{m}\right)^2 +
\left(\frac{T}{m}\right)^{1/2}\right].
\end{equation}


\section{Numerical Results}

Using the above equations we have calculated matter and CMB power spectra
for two different dark matter models:  HDM ($m=10$ eV)
and warm dark matter ($m = 1$ keV) over
a range of scattering cross sections.
In practice we have incorporated the equations
into the CMBFAST code developed by Seljak and Zaldarriaga \cite{SZ96}.
All the models were done assuming that $\Omega_X=0.95$ and
$\Omega_B=0.05$, $H_0 = 50 \, {\rm km \, s}^{-1} \, {\rm Mpc}^{-1}$. The conclusions
are unchanged if a $\Lambda$CDM model is assumed, since our purpose here is only to show how
self-interactions change the power spectra.
Fig.~1 shows the matter power spectrum in terms of the quantity
\begin{equation}
\Delta^2(k) \equiv \frac{k^3 P(k)}{2\pi^2},
\end{equation}
for our two different cases. In both cases, the power spectrum
cut-off is pushed towards higher $k$ if self-interaction is assumed.
The HDM ($m = 10$ eV) results are in agreement
with the results of Atrio-Barandela and Davidson \cite{AD},
for $k$ smaller than the cut-off scale. At small scales, their results are
somewhat different from ours, probably because of an erroneous
term in their perturbation equations (as explained in the appendix).

For our choice of particle masses, the dividing line between the non-interacting
and strongly interacting regimes is roughly at
\begin{equation}
\sigma_0 \simeq 10^{-36} \, {\rm cm}^{2}.
\end{equation}
\begin{figure}[h]
\begin{center}
\epsfysize=12truecm\epsfbox{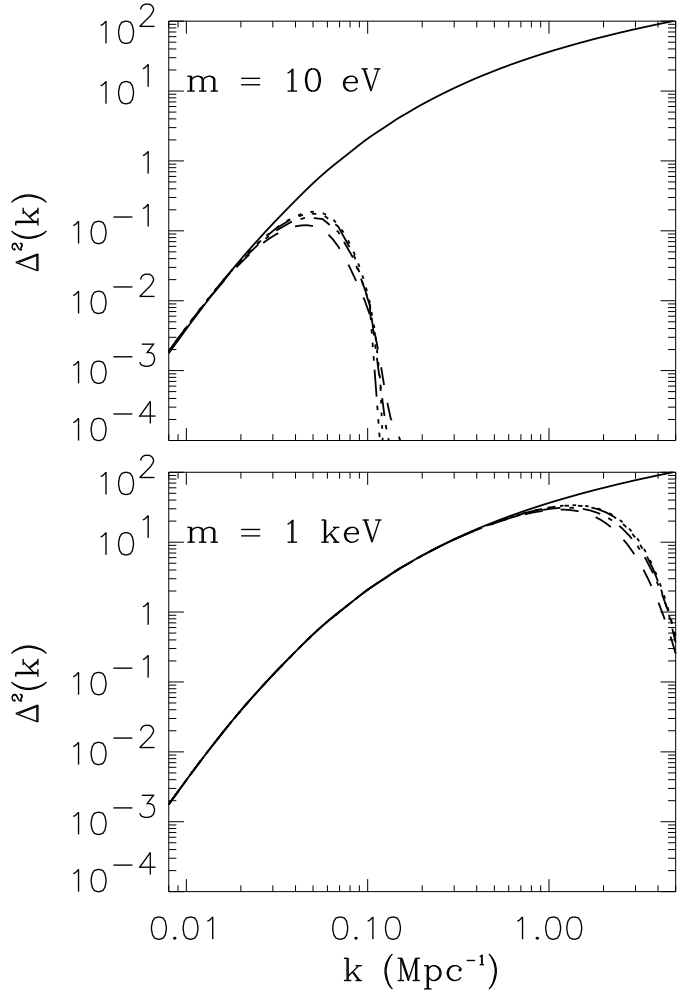}
\vspace{0truecm}
\end{center}
\vspace*{1cm}
\caption{Matter power spectra for two different dark matter particle masses.
The dashed line is for no self-interaction, the triple-dot-dashed is for
$\sigma_0 = 1.2 \times 10^{-38} \, {\rm cm}^2$, the dot-dashed for
$\sigma_0 = 8 \times 10^{-36} \, {\rm cm}^2$, and the dotted is assuming complete
pressure equilibrium. For reference we have plotted the spectrum for standard
cold dark matter (full line).}
\label{fig1}
\end{figure}
Note that this is much lower than the cross section which is needed to explain
structure on galactic scales in the self-interacting cold dark matter model.
In that case, the dividing line is closer to $10^{-23} \, {\rm cm}^2$.
For the case where the dark matter is hot, self-interactions are not able
to improve the agreement with observations significantly because the power spectrum cut-off
is still at much too large a scale.
As discussed in Ref.~\cite{HD}, warm dark matter provides a good fit to
observations of dwarf galaxies if the power spectrum cut-off is at roughly
$2 h_{50} {\rm Mpc}^{-1}$, corresponding to a mass of 1 keV. However, explaining
the core structure of dark matter halos requires that $m \lesssim 300$ eV \cite{HD},
so that even though the uncertainties involved in determining the best cut-off
scale are as large as a factor two \cite{HD}, the collisionless warm dark 
matter model is inconsistent with observations. 
Our results indicate that it might be possible to lower the warm dark matter particle
mass to this smaller
value and compensate by making the warm dark matter self-interacting, which
decreases the cut-off length scale by about a factor of 1.6 compared to the
non-self-interacting case.
Numerically we find that the $k$ where $\Delta^2(k)$ takes its maximum
value is well approximated by
\begin{equation}
\Delta^2(k)_{\rm max} \simeq
\cases{1.1 \left(\frac{m}{\rm 1 keV}\right)^{3/4} {\rm _Mpc}^{-1}
& collisionless, \cr
1.7 \left(\frac{m}{\rm 1 keV}\right)^{3/4} {\rm Mpc}^{-1}
 & strongly self-interacting.}
\end{equation}
For the collisionless case this corresponds to the free-streaming scale,
whereas in the strongly interacting case it corresponds to the
Jeans scale for a given particle mass.
From this result we conclude that self-interacting warm dark matter
is marginally consistent with the present observational constraints.

For the CMB, the fluctuations are usually expressed in terms of the $C_l$ coefficients,
$C_l = \langle |a_{lm}|^2\rangle$, where
the $a_{lm}$ coefficients are determined in terms of the real angular temperature
fluctuations as $T(\theta,\phi) = \sum_{lm} a_{lm} Y_{lm} (\theta,\phi)$.
Fig.~2 shows the CMB spectra for the same two particle masses.
If the dark matter is hot, the CMB spectrum is changed relative to cold dark matter,
because the DM particles are not completely non-relativistic at recombination.
This gives rise to what is called the early integrated Sachs-Wolfe (ISW) effect.
Self-interactions have very little impact because they only affect scales within
the dark matter sound horizon at recombination. Even for a dark matter mass of
10 eV, this is at too small a scale to have a significant impact. For a dark
matter particle mass of 1 keV, the effects are completely negligible.
Our results for non-self-interacting warm dark matter agree with those of
Burns \cite{burns}; we have extended his results to demonstrate that
the addition of self-interactions to the warm dark matter model also
produces a negligible difference from standard CDM.
\begin{figure}[h]
\begin{center}
\epsfysize=12truecm\epsfbox{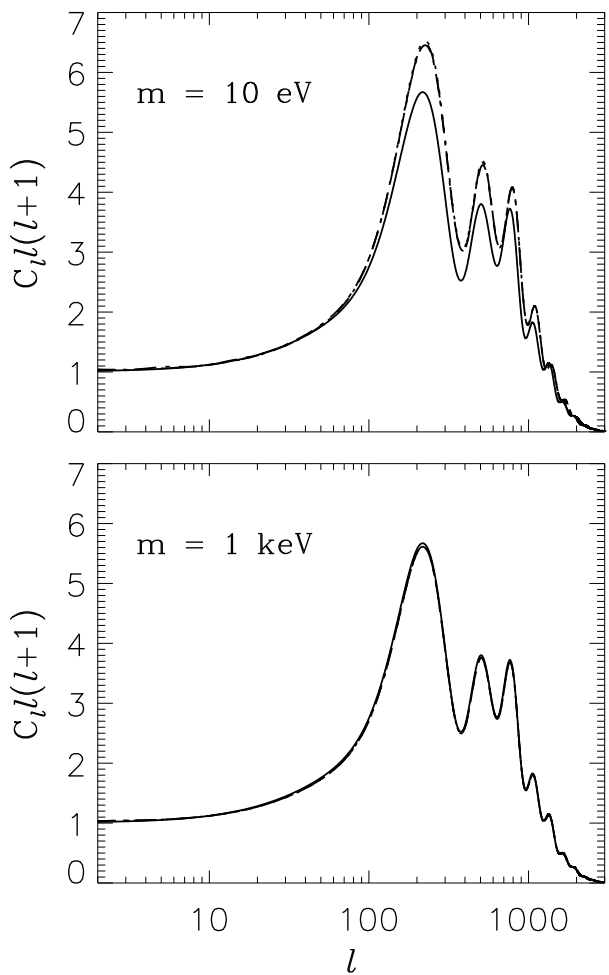}
\vspace{0truecm}
\end{center}
\vspace*{1cm}
\caption{CMB power spectra for the same models as in Fig.~1. The curve labels are
also identical to those in Fig.~1.}
\label{fig2}
\end{figure}


\section{discussion}

We have performed a quantitative calculation of the linear behaviour of
warm dark matter models with possible self interactions.
As expected, power on small scales is generally increased in self-interacting
models because free streaming is suppressed. In collisionless models, power
is suppressed on the free streaming scale, whereas in strongly self-interacting models
the cut-off is at the Jeans scale. 
This increase in the amplitude of the fluctuations
on small scales has the effect of pushing the cut-off
in the power spectrum down to smaller scales by approximately a factor of 1.6.
This may
allow warm dark matter to better fit the dwarf galaxy observations
for masses which are small enough to explain the core structure
of dark matter halos, a result which
could make warm dark
matter a more viable dark matter candidate.

Our CMB results indicate that, like standard warm dark matter, self-interacting warm
dark matter is indistinguishable from standard cold dark matter
in terms of the CMB fluctuation spectrum.  Thus, it is one of the few
variants on the standard model which will not be probed by future CMB experiments.
Any constraints on this model must therefore come from large-scale and galactic
structure considerations.  For instance,
analysis of high-$z$ structure
like damped Ly-$\alpha$ systems might lead to interesting constraints.

Note that 
the cross section for scattering of dark matter particles would have to be
of the order $10^{-36} {\rm cm}^2$ in order to change the matter
power spectrum significantly. This is orders
of magnitude more than the cross sections typical in weak interactions, and
at present there are no obvious candidates for such dark matter particles.
However, it could well be that warm dark matter with relatively strong
self-interactions could be in a mirror sector, in which case there are no
real restrictions \cite{mirror}.


\acknowledgements

SH gratefully acknowledges support from the Carlsberg foundation. 
RJS was supported by the Department of Energy (DE-FG02-91ER40690).
All the 
numerical calculations have been performed using the publicly available 
code CMBFAST developed by Seljak and Zaldarriaga \cite{SZ96}.

\appendix


\section{The boltzmann equation in different asymptotic limits}

\subsection{Large scattering cross sections}

In the limit of very large scattering cross sections, the dark matter is kept in pressure
equilibrium until the present. This is the type of evolution assumed in
Refs.~\cite{CMH92,machacek,LSS}.
In this case the evolution equations read
\begin{eqnarray}
\dot\Psi_0 & = & -k\frac{q}{\epsilon}\Psi_1 + 
\frac{1}{6}\dot h \frac{d \ln f_0}{d \ln q} \\
\dot\Psi_1 & = & k \frac{q}{3 \epsilon} \Psi_0 \\
\Psi_{l \geq 2} & = & 0.
\end{eqnarray}
By performing the appropriate momentum integrations this yields
\begin{eqnarray}
\dot\delta & = & -(1+\omega)\left(\theta+\frac{\dot h}{2}\right)-
3 \frac{\dot a}{a} \left(\frac{\delta P}{\delta \rho} - \omega \right) \delta \\
\dot \theta & = & \frac{\dot a}{a} (1-3 \omega)\theta - \frac{\dot \omega}{1+\omega}
\theta + \frac{\delta P/ \delta \rho}{1+\omega} k^2 \delta
\end{eqnarray}
This equation
is equivalent to Eqs.~(13-14) in Ref.~\cite{LSS}
(when their $\Gamma=\Pi=0$), which are written in gauge invariant form.

\subsection{Large k limit}

At very small scales one may as a first approximation neglect the metric
perturbations. The Boltzmann hierarchy can be truncated by neglecting terms
higher than second order (including $\dot \sigma$), similar to how the Enskog expansion
is performed \cite{relaxation}.
Then the hierarchy equations
when integrated over momentum yield
\begin{eqnarray}
\dot\delta & = & -\frac{4}{3} \theta \nonumber \\
\dot\theta & = & k^2 (\delta/4 - 4 \theta \tau/15).
\label{eq:pert}
\end{eqnarray}
It is interesting to compare our set of equations with Eqs.~(25-26)
of Atrio-Barandela and Davidson (AD) \cite{AD}.
They are almost identical, 
except for the term proportional to $H$ in their equation.
For relativistic particles this term should be zero, as it is in the
above equation.

The term $4 \theta \tau/15$ can be interpreted as a shear viscosity
term, which can in general be written as $\eta \theta/\rho$ \cite{AD}.
Here $\eta$ is the viscosity of the fluid. Using this parametrization
we find that
\begin{equation}
\eta = \frac{4}{15} \rho \tau.
\end{equation}
For a relativistic gas with Boltzmann statistics, $\rho = 3 T n$, so that
\begin{equation}
\eta = \frac{4}{5} T n \tau.
\end{equation}
This expression for the fluid viscosity agrees with what is found
in Ref.~\cite{AD} (their Eq.~(33)).
From Eq.~\ref{eq:pert}, 
one can see that the perturbations oscillate and are damped at
the rate 
\begin{equation}
\Gamma = \frac{2}{15} \tau k^2.
\end{equation}



\begin{references}
\bibitem{peacock}J.~A.~Peacock, ``Cosmological physics'', 
Cambridge University Press (1999).
\bibitem{gross}See for instance M.~Gross {\it et al.},
Mon.\ Not.\ R.\ Astron.\ Soc.\ {\bf 301}, 81 (1998). 
\bibitem{SNIa}S.~Perlmutter {\it et al.}, Astrophys.\ J.\ {\bf 517}, 565 (1999);
P.~M.~garnavich {\it et al.}, Astrophys.\ J.\ {\bf 509}, 74 (1998).
\bibitem{moore}B.~Moore {\it et al.}, astro-ph/9907411 (1999).
\bibitem{ghigna}S.~Ghigna {\it et al.}, astro-ph/9910166 (1999).
\bibitem{NFW96}J.~F.~Navarro, C.~S.~Frenk and S.~D.~M.~White,
Astrophys.\ J.\ {\bf 462}, 563 (1996).
\bibitem{FP94}R.~Flores and J.~R.~Primack, Astrophys.\ J.\ {\bf 427},
L1 (1994).
\bibitem{N99}J.~F.~Navarro, astro-ph/9807084 (1998).
\bibitem{NS99}J.~F.~Navarro and M.~Steinmetz, astro-ph/9908114 (1999).
\bibitem{moore2}B.~Moore {\it et al.}, astro-ph/9903164 (1999).
\bibitem{paolo}See for instance P.~Salucci and A.~Burkert,
astro-ph/0004397, to appear in Astrophys.\ J.\ Lett.\ (2000).
\bibitem{KL99}M.~Kamionkowski and A.~R.~Liddle,
astro-ph/9911103 (1999).
\bibitem{SS88}R.~Schaefer and J.~Silk,
Astrophys.\ J.\ {\bf 332}, 1 (1988).
\bibitem{SLD99}J.~Sommer-Larsen and A.~Dolgov, astro-ph/9912166.
\bibitem{HD}C.~J.~Hogan and J.~J.~Dalcanton, astro-ph/0002330.
\bibitem{SS99}D.~N.~Spergel and P.~J.~Steinhardt, 
astro-ph/9909386 (1999).
\bibitem{H99}S.~Hannestad, astro-ph/9912558.
\bibitem{burkert}A.~Burkert, astro-ph/0002409.
\bibitem{firmani}C.~Firmani {\it et al.}, astro-ph/0002376.
\bibitem{yoshida}N.~Yoshida {\it et al.}, astro-ph/0002362.
\bibitem{MGJPQ}B.~Moore {\it et al.}, astro-ph/0002308.
\bibitem{ostriker}J.~P.~Ostriker, astro-ph/9912548.
\bibitem{ME}J.~Miralda-Escude, astro-ph/0002050.
\bibitem{peebles}P.~J.~E.~Peebles, astro-ph/0002495.
\bibitem{RS}G.~Raffelt and J.~Silk, Phys.\ Lett.\ {\bf B192}, 65 (1987).
\bibitem{AD}F.~Atrio-Barandela and S.~Davidson, Phys.\ Rev.\ D {\bf 55}, 5886
(1997).
\bibitem{CMH92}E.~D.~Carlson, M.~E.~Machanek and L.~J.~Hall,
Astrophys.\ J.\ {\bf 398}, 43 (1992).
\bibitem{machacek}M.~E.~Machacek, Astrophys.\ J.\ {\bf 431}, 41 (1994).
\bibitem{LSS}A.~A.~de Laix, R.~J.~Scherrer and R.~K.~Schaefer,
Astrophys.\ J.\ {\bf 452}, 495 (1995).
\bibitem{MB}C.-P.~Ma and E.~Bertschinger, Astrophys.\ J.\ {\bf 455}, 7 (1995)
\bibitem{SZ96}U.~Seljak and M.~Zaldarriaga, Astrophys. J. {\bf 469}, 437 (1996).  
http://www.sns.ias.edu/$\sim$matiasz/CMBFAST/cmbfast.html
\bibitem{relaxation}See for instance S.~Chapman and T.~G.~Cowling,
``The mathematical theory of non-uniform gases'',
Cambridge University Press (1970); S.~R.~de Groot and P.~Mazur,
``Non-equilibrium thermodynamics'', Horth Holland Publishing Company (1962);
K.~Huang, ``Statistical Mechanics'', John Wiley and Sons (1987). 
\bibitem{burns}S.~D.~Burns, astro-ph/9711304.
\bibitem{mirror}See for instance R.~Foot and R.~R.~Volkas, 
Phys.\ Rev.\ D {\bf 52}, 6595 (1995); R.~N.~Mohapatra and V.~L.~Teplitz,
astro-ph/0001362.
\end{references}
\end{document}